# Thermally Adaptive Surface Microscopy for brain functional imaging

*H.L.M. Robert[1], G. Faini[1], C. Liu[1], N. Rutz[2], A. Aggoun[1], E. Putti[1], J. Garcia-Guirado,[2] F. Del Bene[1], R. Quidant[2], G. Tessier[1], P. Berto[1,3,4,\*]*

[1] Sorbonne Université, Institut de la Vision, Paris, France.
[2] Nanophotonic Systems Laboratory, ETH Zürich, Switzerland.
[3] Université Paris-cité, Paris, France.
[4] Institut Universitaire de France.

E-mail :
pascal.berto@parisdescartes.fr

**Abstract**
Fluorescence microscopes can capture the dynamics of living cells with high spatio-temporal resolution in a single plane. However, monitoring rapid and dim fluorescence fluctuations, *e.g.* induced by neuronal activity in the brain, remains challenging for 3D-distributed emitters due to out-of-focus fluorescence background, a restricted photon budget, and the speed limit of conventional scanning systems. Here, we introduce a Thermally Adaptive Surface strategy, allowing the parallel imaging of 3D-distributed objects at speeds only limited by the camera framerate or photon budget. This microscope add-on leverages on an array of thermally tuneable microlenses that offers low chromatic aberration and high transmission, and can be combined with patterned illumination to provide optical sectioning. We demonstrate its potential *in vivo,* by simultaneously monitoring fast fluorescent dynamics at different depths in the zebrafish larval brain, at 0.5 kHz rate within a 360×360×90 µm$^3$ observable volume.

**Introduction**
Deciphering the dynamics of information flow within neuronal networks is crucial to understand how the brain operates to generate complex behaviours. This requires *in vivo* recordings of neuronal populations in large volumes and distant brain areas with cellular and millisecond resolutions, enabling a comprehensive view of their activity patterns. However, in microscopy, capturing 3D information at high speed remains a challenge when precise spatial selectivity is needed. In practice, classical optics provide no straightforward way to conjugate a flat 2D camera sensor with such complex 3D regions as the brain surface, or neuronal layers. Moreover, most neuronal connections are directed towards layers inside the brain, and investigating transversal cross layer connectivity demands the ability to simultaneously monitor neurons located at different depths.
*In vivo* optical recordings of neuronal activity are commonly achieved via Genetically-Encoded Calcium/Voltage Indicators (GECI/GEVI).[1] When expressed in neurons, these fluorescent proteins can detect and translate changes in calcium ion concentration (a proxy for neuronal activation, for GECIs) or cell membrane potential (a direct measure of neuronal activity, for GEVIs) into variations of their baseline fluorescence. While the last generations of GECIs can reveal fast calcium transients with kinetics down to tens of ms,[2] GEVIs[3] can resolve subthreshold activity, and single action potentials (ms range). In both cases, monitoring short and relatively weak fluorescence bursts requires fast and sensitive acquisition systems. Although the acquisition rates (0.1-1kHz) of modern cameras is not a limiting factor, the restricted photon yield of these indicators demands an optimized collection and a minimization of noises and backgrounds. This can be readily achieved in 2D samples (*e.g.* cultured neurons), but imaging a surface becomes challenging when fluorescence is emitted everywhere in the 3D volume of the brain. Light-sheet,[4] structured illumination,[5] targeted illumination,[6] confocal one-photon[7,8] or two-photon[9–12] microscopes now meet some of these requirements, yet getting information along the third-dimension z at high volume-rate with good optical sectioning remains complex to implement. These needs have fostered the development of various strategies, each with their own trade-offs in terms of speed, SNR, accessible volume as well as resolution, and instrumental complexity.[13–16] They can be classified based on whether the acquisition is performed sequentially or in parallel.

Sequential imaging techniques are based on scanning systems and can achieve high-resolution 3D imaging in living tissues. A first strategy relies on rapid z-axis scans, either by scanning the collection plane and recording full-field 2D images for each depth,[4,17,18] or scanning different targeted illumination positions while integrating fluorescence on the camera.[6,19] Single photon fluorescence is usually combined with confocal approaches,[20,21] while 2-photons systems are mostly used with random-access scanning microscopes[22,23] or scanned multiplexed line(s).[24–27] However, scan-based imaging methods have important limitations: sequential recordings demands a compromise between dwell time (which should be maximized to improve SNR), volume acquisition time (which should be minimized), and the accessible volume (or number of monitored neurons). Increasing the acquisition rate and the scanning speed up to hundreds of Hz is possible but clearly comes at the price of instrumental complexity,[19,21–23] and does not circumvent the issue of photon budget.

A second class of strategies relies on parallel imaging and scanless systems. 1 kHz volumetric single-photon voltage imaging has been demonstrated using a multiplexed excitation and collection microscope, although with a relatively complex system (7 light-sheet beams, 7 cameras and deconvolution post-processing).[28] Light field and multifocus microscopy[29–32] are simpler approaches, able to perform parallel imaging at several depths/angles with acquisition rates limited by the camera and the photon budget. However, the reconstructed signal is generally affected by background fluorescence although interesting hybrid scanning approach have been recently proposed.[33] 3D parallel targeted illumination[34–37] offers an interesting alternative: such microscopes can illuminate selectively and simultaneously several neurons in 3D using a liquid-crystal spatial light modulator (LC-SLM), and collect signals at different depths. However, the diffraction and polarisation losses of LC-SLMs restrain the shaping of the fluorescence light on the collection path, which can be critical in scenarios involving tight photon budgets and dim fluorescence changes.

Moreover, their strong chromaticity forbids the use of a single SLM to shape both the illumination and fluorescence wavelengths. These methods thus employ extended-depth-of-field schemes to project the 3D information over the image plane of the sensor, although this increases complexity and imposes strong limitations on the axial range, resolution, and SNR.

Recently, we proposed a novel wavefront shaping concept coined as *SmartLens* (SL), which exploits electrically-induced microscale thermal phase-shifts.[38–40] As depicted in Figure 1(a-d), these engineered micro-resistors can create a predetermined temperature landscape in a polymer which leads, through the thermo-optical effect, to an adjustable diverging lens. These compact devices are non-birefringent, mostly free of diffraction artifacts and chromatism, and cause minimal optical losses, which is crucial in order to preserve the photon budget, particularly for fluorescence imaging. This makes them excellent candidates to tailor and focus both the illumination and the emitted fluorescence.

Here, we propose a Thermally Adaptive Surface (TAS) microscope, which is capable of simultaneously imaging fluorescent microscopic objects distributed in several planes with a high spatio-temporal resolution. The main component of this microscope is a mere add-on module based on an array of active thermal SmartLenses, the SLs,[38] enabling imaging at independently selected depths in multiple subregions of the FOV. This is illustrated in Fig. 1 (e-g), where 5 SLs are activated to image green fluorescent neurons in the tail of a zebrafish, at 5 different depths below the focusing plane of the objective. The TAS microscope does not allow the single-shot tomographic imaging of a volume, *i.e.* the acquisition of several regions superimposed along the propagation axis in a single acquisition. However, the SL arrays provide a simple, efficient way to tailor the 3D shape of both the illumination and collection surfaces, owing to their good achromaticity. As such, they allow the concurrent fluorescence monitoring of regions located at different z-positions in a single camera plane, thus optimizing the photon budget and acquisition time along an arbitrary, freeform surface. We show that, associated with targeted or structured illumination, TAS microscopy can simultaneously monitor objects distributed in a 3D volume, with reduced fluorescence background. We demonstrate that this scanless system is particularly suited for fast neuronal activity monitoring as it enables camera-framerate-limited, cellular-resolution imaging in multiple regions (up to 25 here) across ten different planes and within a volume of $360\times360\times90$ µm$^3$. We validate this approach *in vivo* using one-photon microscopy, capturing fast fluorescence dynamics at a frequency of 0.5 kHz within the volume of zebrafish larval brains.

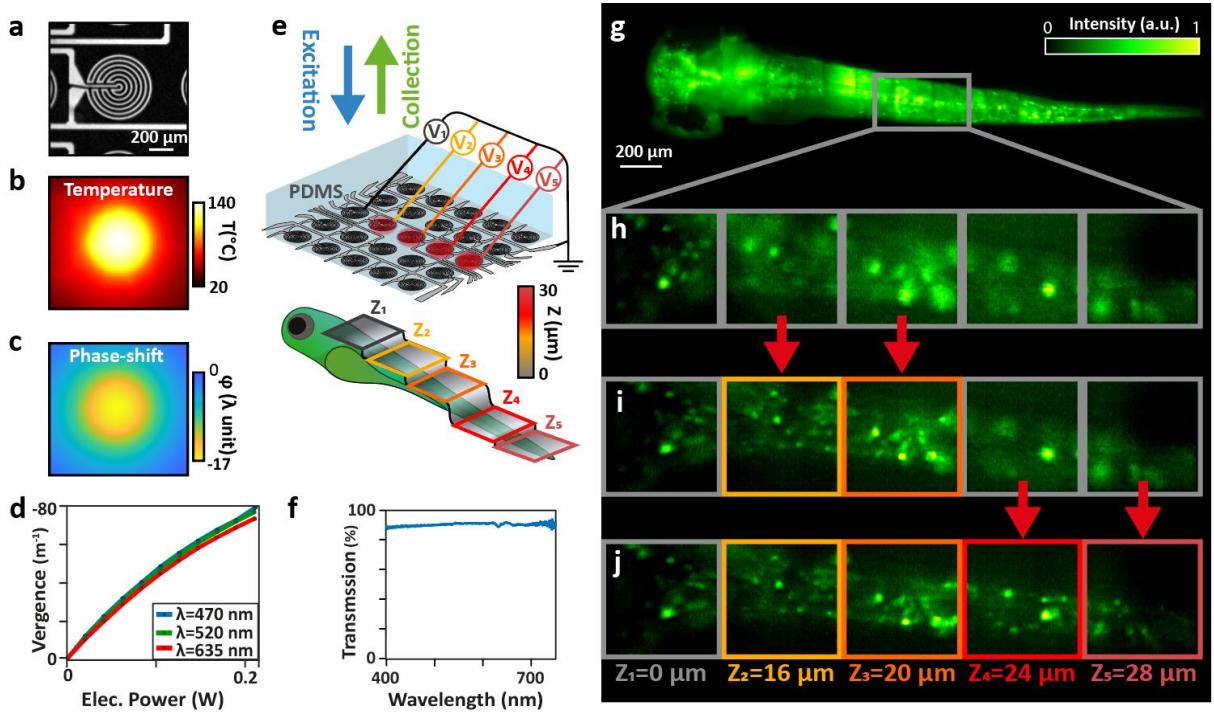

**Fig. 1: Principle of the Thermally-Adaptive-Surface (TAS) microscope. a** Optical image (acquired in reflection) of a single thermal micro-actuator "*SmartLens*" (SL). **b** Measured temperature map within the SL when applying an electrical power ($P_e = 0.2$ W) leading to a refractive index modulation in the thermo-responsive polymer (PDMS). **c** Resulting phase-shift measured with a high-resolution wavefront sensor (at λ=532 nm), showing that a single SL locally acts as a microlens. **d** Vergence of a single SL as a function of supplied power, for 3 different wavelengths, showing low chromaticity. **e** Principle of the TAS microscope based on a SL array. **f** Transmission spectrum of a single SL, showing low losses. **g** Composite fluorescence image of a zebrafish larva expressing the GCaMP8m calcium indicator. **h-j** Fluorescence images in the tail of a zebrafish. The coloured squares outline the active areas (≈70×70 µm² in the sample space) associated to each SL. While only the left part of the image ($Z_1$) is in focus in the fluorescence images when all SLs are off (h), powering on two SLs (i) or four (j) allows to access multiple image planes ($Z_1$ to $Z_5$).

## Results

### *SmartLenses* arrays for adaptive surface imaging

The key component of the TAS microscope is its SL array, a custom-made electro-thermally-adaptive microlens array. It is composed of a 5×5 independent actuators (SL),[38] consisting of transparent, conductive Indium Tin Oxyde (ITO) microwires fabricated using classical UV lithography (see Methods). In the active part of the thermal microlens, the wire width is reduced to locally increase the resistance. Each spiral-shaped actuator has an outer diameter of 540 µm, as illustrated in Fig. 1a (a full image of the array is shown in Supplementary Figure S1). A 500 µm-thick polydimethylsiloxane (PDMS) layer used as thermo-optical material is deposited on top of the resistors. Applying voltage to a SL generates heat through Joule effect, particularly in the narrow-wire/high-resistance-regions, resulting in a local temperature increase (Figure 1b). This reduces the refractive index of the heated regions since the thermo-optical coefficient of PDMS ($[dn/dT]_0 = -4.5 \times 10^{-4} K^{-1}$) is negative. Therefore, an optical plane wave undergoes a smooth, negative phase-shift (Fig. 1c), leading to a local divergence of the optical beams. Each SL can be independently activated with a maximum power of 0.2W, which corresponds to a maximum temperature of 140°C as shown in Figure 1b (details on power consumption are in Supplementary section S2). The safety margin with respect to the 250 °C ceiling temperature of PDMS is sufficient to avoid damage.[41] Varying the power from 0 to 0.2 W yields a tuneable diverging microlens with a vergence $V_{SL}$ ranging from 0 to -80 m$^{-1}$, corresponding to a focal distance range [-∞, -13 mm], and a maximum *f*-number of 24 (see Figure 1d).

This wavefront shaping strategy offers important advantages over standard liquid crystal SLM (LC-SLM): thermal shaping generates smooth, phase-wrapping-free profiles and does not generate diffractive chromatic aberrations. SLs are therefore quasi-achromatic devices as shown in Fig. 1d. Moreover, they are polarization-insensitive, thus avoiding the 50% loss incurred when using LC-SLMs with non-polarized light (including fluorescence). On top of that, SLs are based on 50 nm-thick indium tin oxide (ITO) wires allowing a high transmittance (T ≈90%) and low diffractive losses (typ. 5%) in the visible range as shown in Fig. 1f. SL arrays are thus particularly adapted to low yield, unpolarized, and intrinsically polychromatic (excitation and emission) fluorescence imaging.

## Description of the TAS microscope

The principle of the TAS microscope is depicted in Figure 2a. A collimated excitation laser (λ=473 nm) illuminates the surface of a digital micromirror device (DMD). The DMD is conjugated to an Intermediate image Plane (IP) through an afocal system and reimaged in the (**x,y**) object plane of a microscope objective (Olympus ×20, NA= 0.5). The one-photon epifluorescence signal is then collected within a Field of View (FOV) of 360 × 360 µm² by the same objective and is imaged on a camera (Kinetix22 SCMOS) after being filtered by a long pass dichroic mirror (cut-off at λ=500 nm) and a bandpass filter (520±28 nm). The DMD, conjugated with the (**x,y**) sample plane, allows to pattern the illumination of the sample. It enables targeted or structured illumination, which enhances the signal-to-background ratio, provides optical sectioning, and reduces photobleaching.

To locally control the axial focusing of the object plane, the SL array is placed a short distance δ away from the Intermediate Plane (IP) conjugated with the object plane of the objective (details on the SL array positioning are provided in Supplementary section S3). Each active SL area defines the subFOV in which it can induce a defocus in the sample space, FOV size/ number of SLs, typ. 70×70 µm² (details on the optimal SL pupil diameter are provided in Supplementary section S4). Let us consider a given point source on the DMD, imaged at the IP within a given subFOV. The activation of the corresponding diverging thermal microlens will shift its image away from the IP by a distance $z_{shift}$ (red dashed line in the central inset of Fig. 2a). In the sample space, the image of this point source will therefore be shifted by a distance $z'_{shift}$ (bottom inset in Fig. 2a), to excite a targeted fluorescent point source (*e.g.* a neuron). This focus shift can be expressed (see Supplementary section S3 for derivation) as:

$$z'_{shift} = -\frac{V_{SL}\delta^2}{1+V_{SL}\delta}M_L , \qquad (1)$$

with $M_L$ the longitudinal magnification between $z_{shift}$ and $z'_{shift}$, and $V_{SL}$ the vergence of the activated SL. It is important to note that the fluorescence emitted by the targeted point-source and collected by the objective passes through the same thermal lens which is, as discussed above, relatively achromatic (see Fig. 1d). Therefore, the fluorescent point-source remains conjugated to the image/camera plane whatever the defocus applied by the SL (see top inset in Fig. 2a).

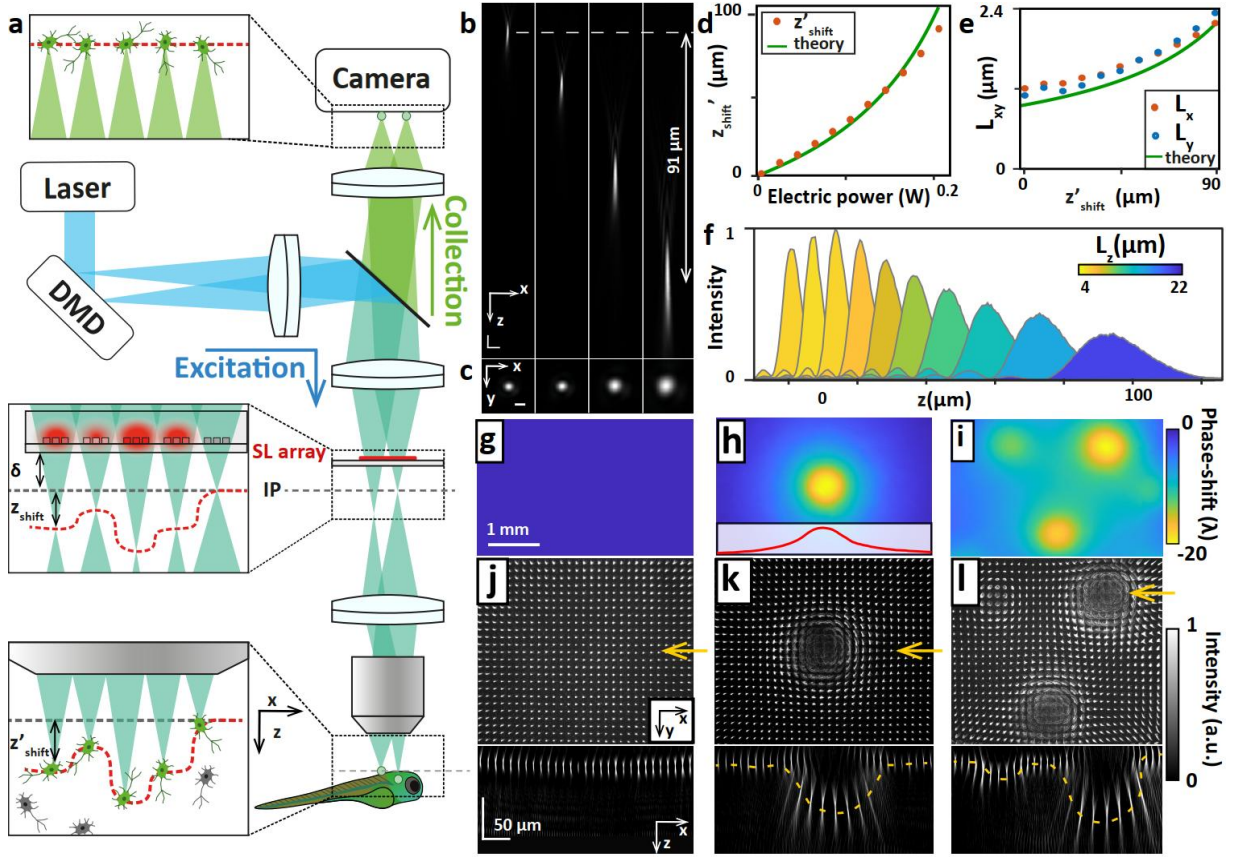

**Fig. 2: Experimental setup and characterization of the TAS microscope. a** Optical design of the TAS microscope. IP=Intermediate image Plane. Middle inset: focus shift $z_{shift}$ (red dotted line) induced by the SL array near the Intermediate Plane IP (grey dotted line). Bottom Inset: focus shift $z'_{shift}$ induced by the SL array in the sample space. Top inset: each subFOV remains conjugated to the camera plane regardless of the applied defocus since both the excitation and the collected fluorescence pass through the quasi-achromatic SLs. (**b-c**) Images of a single PSF (excitation) for powers $P_e = 0, 63, 147$ and $210$ mW dissipated in the SL, along the (x,z) (b) and (x,y) (c) planes. Scale bars in (b): 5 µm, in (c): 1 µm. **d** Focus change in the sample space $z'_{shift}$ and subFOV magnification as a function of the electric power. **e** Variation of the lateral full width at half maximum (FWHM) of the PSF beam, $L_x$ and $L_y$, measured in the sample space as a function of the defocus $z'_{shift}$. **f** Intensity cross-sections along the z-axis as a function of the distance z. The colours indicate the corresponding axial FWHM, $L_z$. **g-i** Images of the thermal phase-shift induced by the SL array when (g) all the SLs are off, (h) the central SL is on ($P_e=0.18$ W; inset: 1D profile) and (i) 4 actuators are on (from left to right: $P_e=0.1, 0.16, 0.18$ and $0.09$ mW). **j-l** Experimental images of the excitation light, structured in a 2D grid of 11 µm-spaced dots in the sample space by the DMD, along the (x,y) plane (top) and the (x,z) plane (bottom), in the same conditions as in g-i. The yellow arrows represent the position of the profiles plotted in the bottom images. The dashed lines indicate the imaging surface.

The same applies within any of the 5×5 subFOVs corresponding to each SL: the camera is always conjugated to the DMD plane, regardless of the displacement $z'_{shift}$ of both the illumination and fluorescence imaging depths in the sample. Activating one or several of the 25 SLs in the array therefore allows an adaptive adjustment of both the excited and imaged surfaces within the sample volume.

## Characterization of the microscope

To characterize the 3D size and the shift of the excitation PSF, we implemented a forward imaging path employing an objective with a higher NA than the objective of the TAS microscope (full optical design available in Supplementary Figure S5). Since the DMD pixels correspond to sub-diffraction regions in the sample space, activating a single pixel amounts to using a point source and allows direct PSF characterization. 3D PSF shapes were measured by axially scanning the collection objective for various heating powers applied to the SL facing the illuminated pixel (Figures 2b-f and full data shown in Supplementary Figure S6). In this configuration, the TAS microscope provides a focusing range of 91 µm. Note that the measured focus displacement in the sample space $z'_{shift}$ is accurately described by Eq. 1 (green line in Fig. 2d), although this simple model only considers the SL as a thin lens rather than a continuous 3D refractive gradient.

As shown in Figures 2 b, c, e, f, the axial and transverse full width at half maximum (FWHM) of the beam, respectively $L_z$=4-24 µm and $L_{xy}$ =1.2-2.4 µm, both increase with the focusing depth. In imaging mode, these translate into a slight reduction of the transverse resolution and an increased depth of field (more details in Supplementary section S7). Note that this effect is not caused by SL aberration but rather by the positioning of the SL array. In most standard configurations, a single tunable lens is positioned in the pupil plane to control a global defocus. Here, to enable independent control of different subFOVs, the SL array is located *near* a conjugated plane. In this non-telecentric configuration, activating a SL therefore reduces the effective NA and causes an underfilling of the back pupil of the objective, leading to an increase in the axial and lateral extension of the excitation PSF. Taking this into account, i.e. dividing the accessible $z'_{shift}$-range (91 µm) by the PSF axial extensions ($L_z$) at each depth, one can consider that up to 10 planes can be unambiguously distinguished (see Fig. 2f). It is worth mentioning that, for moderate $z'_{shift}$ distances compared to the objective focal length (here $z'_{shift} \approx$100 µm and $f'_{obj}$=9 mm), the number of accessible planes $N_{planes}$ can be simply expressed as (see Supplementary section S8 for derivation):

$$N_{planes} = - \frac{D_{SL}^2 V_{SL\,max}}{2\pi\lambda}, \qquad (2)$$

with $D_{SL}$ the effective diameter of the SL (details in Supplementary section 4), $\lambda$ the excitation wavelength, and $V_{SL\,max}$ the SL vergence range. Using equation (2) and the TAS microscope parameters, we have $N_{planes}$=11, which is close to the experimental value. Moreover, we see from equation (2) that, for a given SL diameter, the number of distinguishable planes is independent of the NA and the magnification of the objective.

To visualize the adaptive excitation surface when activating SLs (Fig. 2g-i) and to characterize the PSF distortions over the entire FOV, we displayed a grid of 33×33 evenly-spaced points on the DMD and measured the resulting 3D intensity distribution in the sample space (see Fig. 2j-l and 3D animations in Supplementary video 1). Although affected by some aberrations (field curvature is typical of non-telecentric systems[42]), the local diverging lens effect is clearly visible in the grid distortion as shown in Figure 2h and k. As shown by the xz projection in Fig. 2.l, PSFs located at three distinct depths, respectively $z'_{shift}$=0, 20 and 65 µm are clearly observed, clearly showing that the TAS microscope can simultaneously shift the local focusing depths within multiple subFOVs. Complementary studies of the local magnification changes within subFOVs can be found in Supplementary section S9.

Fast and simultaneous monitoring of 3D neuronal activity.

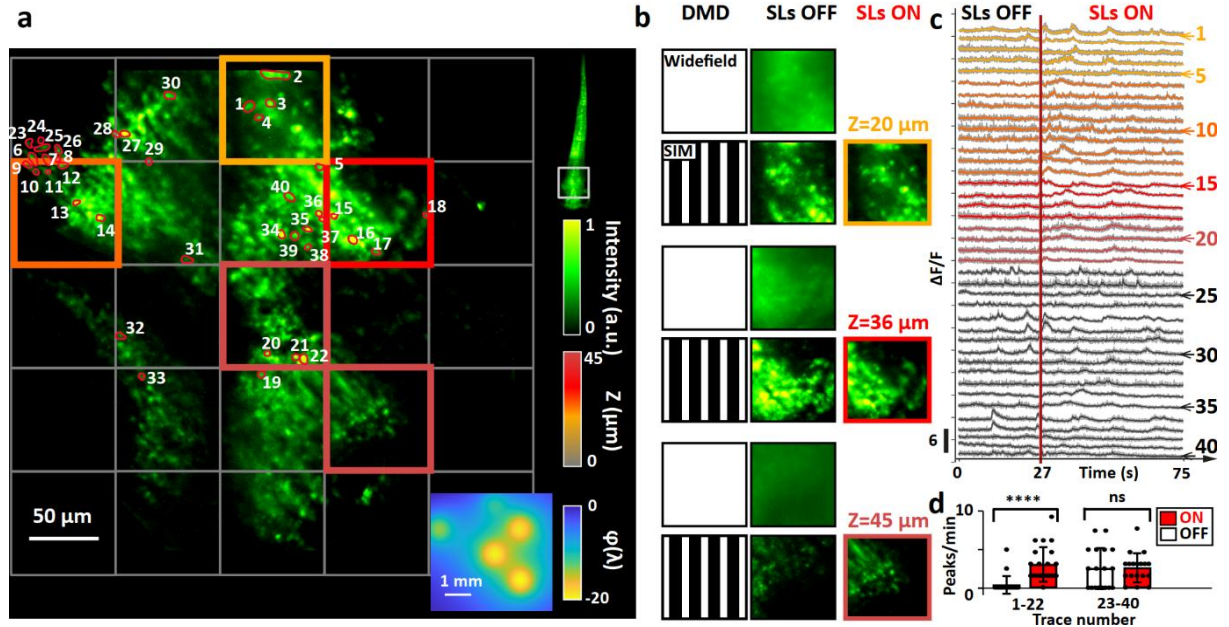

**Fig. 3: Monitoring zebrafish neuronal activity using Structured Illumination Microscopy (SIM). a** Reconstructed SIM image of neurons in the zebrafish brain. The grid represents the SL array, and the colours of the squares correspond to the defocusing values $z'_{shift}$. Chosen neurons (red circles) are numbered, and the corresponding relative fluorescence intensities are shown in (c). Bottom inset: Phase image of the activated SL array. Top right vertical inset: composite fluorescence image of a zebrafish larva. **b** Selection of three subFOVs using two illumination configurations: standard full-field illumination (first rows) and structured illumination (second rows); using SLs in the off (first column) and on states (second column). **c** Relative variations of the fluorescence intensity $\Delta F/F$ measured in each of the 40 regions numbered in (a). Neuron activity is measured as a fluorescence increase in neurons expressing GCaMP8m. Light grey: $\Delta F/F$ signal after SIM demodulation (raw images acquired at 150 Hz). In colours, rolling average of 10 consecutive SIM measurements. The vertical red line indicates the instant (t = 27 s) at which SLs are turned on. Each curve was recorded in regions of identical colours in (a), corresponding to $z'_{shift}$= 20 µm, 30 µm, 36 µm and 45 µm defocusing. **d** Box chart of detected peak frequency for traces 1-22 (activated SL regions) and 23-40 (non-activated regions) when the SLs are off (t<27 s) and on (t>27 s). Dots, boxes, lines represent respectively the measured, averaged and standard deviation of the peak frequency for each neuron. **** corresponds to p-value≤$10^{-4}$, "ns" corresponds to non-statistically significant data (p-value>$5.10^{-2}$). The number of detectable events clearly increases where and when SLs are activated (traces 1-22), and is unchanged where they are not (23-40).

To demonstrate the capabilities of the TAS microscope, we performed high-speed, 3D fluorescence measurements in a zebrafish larval brain expressing the GCaMP8m sensor[2]. GCamp8m is a calcium indicator with rapid kinetics and high sensitivity to neuronal activity *i.e* its fluorescence rapidly increases in response to changes in calcium ions concentration. We monitored the neuronal activity of multiple GCaMP8m-expressing neurons, by measuring multiple spontaneous calcium transients simultaneously and at different depths. Since neurons are distributed in the entire volume of the brain, wide-field illumination comes with important background fluorescence, which can be efficiently reduced using Structured illumination microscopy (SIM)[5] or targeted illumination[6]. Here, their applicability strongly relies on the broadband operability of our system, since both the excitation, patterned by the DMD, and the collected fluorescence pass through the same SL.

SIM relies on the excitation of the sample by a known structured pattern (usually stripes) to downshift the spatial frequencies and access sub-diffraction information,[43] but also to reject out-of-plane light.[5] Here, we use SIM for the latter optical sectioning effect, thus rejecting the unwanted fluorescence. The process consists in a sequence of 3 binary-striped patterns which are sequentially swept laterally by 1/3 of the stripe period, and synchronized to the acquisition. After demodulation, this provides a strong reduction of the fluorescence background.

Figure 3a shows a SIM image of fluorescent neurons expressing the calcium indicator GCaMP8m in the brain of a zebrafish larva (full video in Supplementary video 2). The background rejection is clearly visible in Fig. 3b when comparing subFOVs acquired in wide-field and using a SIM illumination. We recorded the spontaneous activity of a total of 40 neurons simultaneously at different depths and measured the corresponding relative fluorescence variations $\Delta F/F$. Here, 5 thermal lenses are activated (coloured squares in Fig. 3a) to simultaneously monitor neurons located at 5 different depths $z'_{shift}$ = 0, 20, 30, 36, and 45 µm. Associated with SIM, SL activation provides an improvement of the focus within the chosen regions, as shown in the images in Fig 3b. This is particularly visible in the relative fluorescence variations $\Delta F/F$ traces (Fig. 3c): the activity (peaks rate) remains relatively constant in neurons located in regions where SLs are not activated (in black, neurons #23-40), showing either a steady activity in neurons which are in focus (*e.g.* #30), or almost no activity in neurons which are out of focus (*e.g.* #31). In contrast, regions where SLs are activated (coloured squares in Fig.3a, and coloured traces #1-22 in Fig.3c), present a sharp increase in the recorded activity after activation of the SLs (t > 27 s, vertical red line): while very few peaks are detected for t < 27 s (SL off), clear fluorescence peaks attributed to neuronal activity are identified for t > 27 s (SL on), demonstrating an improved focusing.

To quantify this gain in calcium transients' detection, we compared the number of observed peaks per minute (Fig. 3d) without (black) or with (red) SL activation. A significant increase in peak rate is visible (p-value<$10^{-4}$) for neurons # 1-22 when the corresponding SLs are activated, whereas this rate is relatively constant in regions # 23-40 where the thermal microlenses remain off (similar box charts recorded in all regions can be found in Supplementary Figure S10). These results clearly indicate that when SIM illumination is used to provide optical sectioning, the activation of SLs allows to recover calcium transients associated to specific depths of interest.

While the SIM approach associated to thermal microlens refocusing drastically reduces background noise and improves the detection of neuronal activity, it clearly reduces the acquisition frame rate by a factor related to the number of images N used in the reconstruction (here N=3). This limitation can be overcome using a targeted (or multi-spot) illumination modality[6] in which only specific areas are illuminated with the DMD. This offers several advantages: the fluorescence background is minimized and the photobleaching is reduced since the energy is only deposited in region of interest. Importantly, this strategy does not involve any post-processing step (as opposed to the above SIM approach) and allows to work at maximum camera framerate. This approach is well adapted to selectively monitor a discrete number of individual neurons inside dense and interconnected cellular populations, ensuring fast activity detection with single-cell resolution.

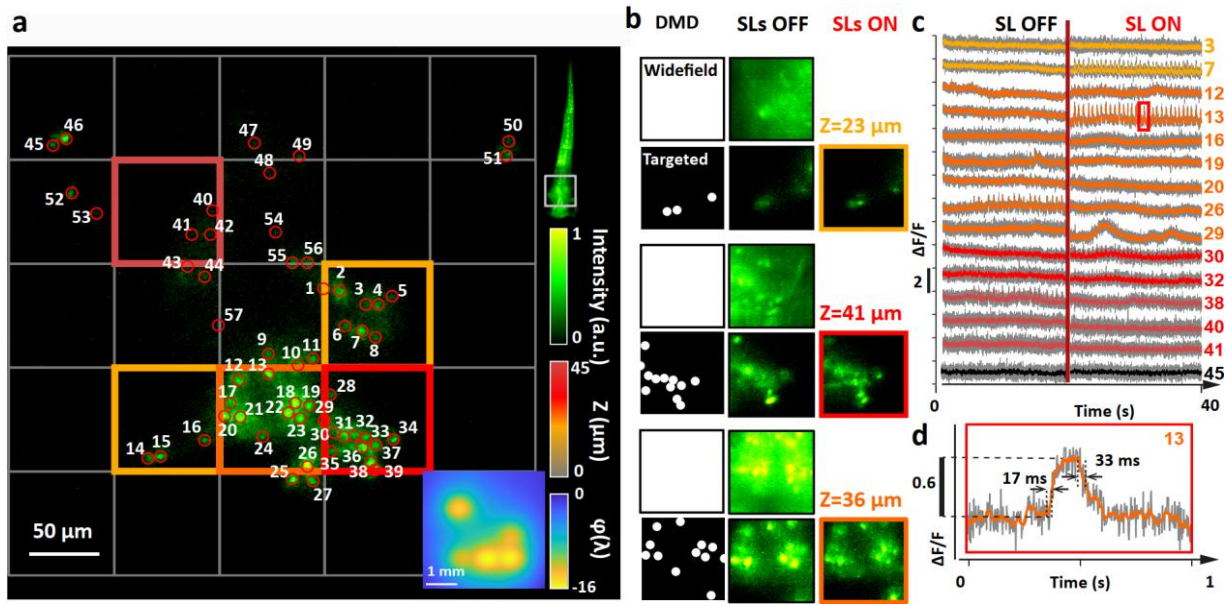

**Fig. 4: Monitoring of zebrafish neuronal activity using targeted (multi-spot) illumination.**
**a** Image of neurons within a zebrafish brain, where 57 neurons are identified, targeted and numbered (# 1-57). Bottom inset: phase image of the thermally activated SL array. Right inset: composite fluorescence image of a zebrafish larva. **b** Selection of three subFOVs illuminated in full field (first rows), or with targeted illumination using 10 µm diameter discs corresponding to the 57 identified neurons (second rows). In the orange-red subFOVs, the SL is either in the off (first column, t < 20 s) or on (second column, t > 20 s) states **c** Relative fluorescence variations $\Delta F/F$ in 15 selected regions, as numbered in (a), before and after SL activation, at t = 20 s (red vertical line). **d** Detail of the fast kinetics observed in neuron 13. In c) and d), the grey curves represent the raw 0.5 kHz-rate recording, and the red/orange/black curves are rolling averages over 10 frames.

Figure 4a shows a zebrafish brain in which the fluorescence of multiple neurons was captured simultaneously at 0.5 kHz rate (full video in Supplementary video 3). Again, switching on the SLs enables the targeting of out-of-focus neurons of interest. The improvement in optical sectioning and refocusing achieved through targeted illumination and SL activation within 3 subFOVs is highlighted in Fig. 4b. Fig. 4c shows the recorded trace of 15 selected neurons (full set of traces in Supplementary Figure S11) located at 5 different depths $z'_{shift}$ =0, 23, 36 ,41 and 45 µm after t= 20 s. Some neurons (# 7, 12, 13 and 29) do not display any observable dynamics before the corresponding SLs are switched on. Conversely, other cells (# 19, 26 and 38) display observable peaks while the SLs are off (t<20 s, $z'_{shift}$=0 µm) but none when the SLs are activated ($z_{shift}'=36$ µm for traces 19 and 26 and $z'_{shift}=45$ µm for trace 38), confirming the axial discrimination capability of the TAS microscope. Note that the high recording speed (0.5 kHz) allows the detection of fast fluorescent signals, with rise and decay times in the 10 ms -range (Fig.4d), highlighting the importance of a full parallel detection to maximize both speed and signal.

## Discussion

In this work, we proposed a novel strategy for the parallel imaging of fluorescent objects distributed in a 3D volume. This scanless method leverages on the thermo-optical effect to independently control the focus in several regions of the FOV. The low chromaticity and broadband operability of this refractive-based approach working in transmission allows to shape both the illuminated and imaged surfaces. Importantly, this microscope is well suited to the study of complex cellular distributions and processes in their natural, non-planar 3D geometries (*e.g.* in the brain). In the context of brain activity recordings, this approach enables the simultaneous monitoring of fast neuronal dynamics from specific neuronal populations or regions of interest, at different depths, with cellular resolution. We demonstrated the capabilities of the system *in vivo* by simultaneously monitoring brain activity in multiple neurons located at different depths in the zebrafish larva, with reduced background, and at a speed limited only by the camera frame rate and the fluorophore photon budget. We show that this system allows focusing in up to 10 (longitudinal) planes and 25 (lateral) regions of interest within a 360×360×90 µm³ volume, thus enabling the monitoring of fast fluorescent signals at a framerate of 0.5 kHz, but the concept is readily extendable to more regions and /or larger volumes and /or higher number of planes. Indeed, as demonstrated in Supplementary section S8, the number of accessible planes is independent of the microscope objective used, and scales linearly with the vergence range of the SL. This range could clearly be extended using materials with larger thermo-optical coefficients[44].

The proposed implementation shares certain similarities with scanless approaches used in optogenetics, which employ LC-SLMs to photostimulate in parallel multiple neurons in 3D.[45] However, fluorescence imaging requires wavefront modulation on both the excitation and collection paths, which requires locating the wavefront shaper (*i.e.*, the SL array) in a conjugate plane rather than in the pupil plane. If attempted with a standard LC-SLM (or LC arrays of microlenses), this would induce crippling diffractive chromatic aberrations as well as diffraction (typ. 10-40%) and polarization losses (typ 50%), which are detrimental to high SNR, high speed neuron imaging. In contrast, broadband, polarization-insensitive, refractive SLs offer high transmission (~90%), which could be further improved with the application of antireflection coatings.

While integrating reflective optics to a microscope requires a folding of the optical path, which notably increases bulk and complexity, these refractive systems operating in transmission mode can be easily integrated to any fluorescence microscope. In much the same way as tunable lenses, SL arrays can act as straightforward add-ons to SIM, HiLO or fast-scan microscopes, while enabling parallel access to three-dimensional information. In association with targeted illumination, they also offer a "what you see is what you get" approach. This allows direct 3D imaging free from processing noise (*e.g.* deconvolution noise) which is important in the low-photon regime. Note that a confocal scheme could be implemented, *e.g.* by using the DMD not only in the excitation path but also in the collection path, such that selective excitation pixels also serve as pinholes to block out-of-focus fluorescence.[7] The fabrication of the SL arrays which are central to multiplane TAS microscopy only requires a simple UV lithography step, making them particularly cost-effective. As such, these TAS modules have a strong dissemination potential.

We should however point out that this adaptive surface strategy is not genuinely tomographic or volumetric imaging: while objects located at different depths along the same longitudinal path can be selectively imaged by tuning the corresponding SL, they cannot be monitored simultaneously. This issue could be partially tackled by improving the response time of the microlenses (currently $\tau \approx 500$ ms) to allow fast commutation between planes. Previous works[38,46] have indeed shown that the SL response time, which is driven by thermal diffusion, can be reduced down to the ms range or below by decreasing the heat source size, the thickness of the thermo-optical medium (here PDMS) or by using overdrive strategies. Fast SL arrays should be able to provide a parallelized scanning of multiple SubFOVs at several depths of interest, thus significantly accelerating volumetric measurements as compared to global scans.

The ability of our system to monitor dim fluorescence changes, in the kHz range or above, is essential to 3D voltage imaging experiments,[3] in order to detect both subthreshold neuronal events and single action potentials. Moreover, combining SL arrays with scanless[12] (*e.g.* based on temporal focusing) or fast-scanning multiphoton[9–11] microscopy approaches should give access to deep, 3D functional imaging with enhanced background rejection, which is crucial in scattering tissues.[47,48] Transmission-mode SL arrays can also be easily engineered to reach sub-millimeter sizes, potentially allowing their integration into miniaturized microscopes or endoscopes. Finally, the broadband capability of this system opens up opportunities for 3D all-optical studies, *i.e* dual-wavelength applications for imaging and optogenetic photostimulation. We demonstrated that the current TAS microscope scheme is well adapted to neuron functional imaging, but it should also emerge as a valuable approach in a wide range of scenarios requiring the rapid detection of optical signals from non-planar structures across various scales, which are common issues in biology. This might include *e.g.* macro-scale brain regions monitoring,[49] blood flow monitoring in meandering blood vessels,[50] or the tracking of calcium, voltage or neurotransmitter signaling along axons and dendritic arbors.[51]

## Methods

Zebrafish preparation

All procedures involving animals were conducted in accordance with European and French guidelines with protocols approved by the committee on ethics of animal experimentation of Sorbonne University (protocol ID: APAFIS# 21323). The fish were maintained at 28.5 °C on a 14 h light/10 h dark cycle in the animal facilities of the Vision Institute.
Briefly, we generated the *Xla.Tubb2-hsp70-ubc:H2B-jGCaMP8m* plasmid by using the Gibson Assembly Cloning Kit (New England Biolabs). The plasmid was co-injected in one-cell stage zebrafish embryos (casper strain)[52] with tol2 transposase mRNA (25ng/uL) and F0 fish were screened for green fluorescence to generate the F1 line. For Fig.1 and 3, we used the resulting stable line. For Fig.4, we used larvae transiently expressing the plasmid. The day of experiment, 4-6 days post fertilization (dpf) zebrafish larvae were mounted in a petri dish using 2% low melting point agarose.

Smartlens Fabrication

The SLs were fabricated using standard UV photolithography techniques. A vapor-phase deposition of an hexamethyldisilazane (HDMS) monolayer was performed on a pre-cleaned 4-inch glass substrate coated with an ITO layer exhibiting a sheet resistivity of 40 Ω/sq at 50 nm thickness (CEC040S, PGO-Online GmbH). Subsequently, a 0.7 µm thick layer of photoresist (AZECI3007) was applied via spin coating. The coated substrate was soft-baked at 110 °C for 60 s, followed by optical lithography exposure for 10 s (Süss MA6). Post-exposure bake was performed again at 110 °C for 60 s. Then, the sample was developed in a 1:4 parts dilution of AZ 400 K with deionized water for 9 s. This was followed by a hard bake at 120 °C for 120 s and subsequent oxygen plasma at 200 W for 30 s prior to the etching step. The ITO layer was etched chemically with HCl(35%):HNO3(65%):H2O in a 1:0.08:1 volume ratio. The etching process was carried out at a controlled temperature of 33±2°C for 140 s, followed by thorough rinsing with deionized water. The residual photoresist was stripped using acetone in an ultrasonic bath, and the substrate was subsequently rinsed with isopropanol (IPA) to ensure a clean surface. Next, the wafer was prepared for dicing. A 1.2 um thick layer of photoresist (AZ1512) was used for protection. Once diced, the photoresist is striped with acetone and rinsed with IPA, then the chips are ready for the next step. A 500 µm-thick PDMS membrane was prepared via drop-casting on a silane-coated silicon wafer using the Sylgard 184 mixed at the standard 1:10 curing agent to pre-polymer parts. After curing, the membrane was precisely cut and adhered to the patterned ITO surface by applying a small drop of uncured PDMS as an adhesive layer. The assembly was baked in a convection oven at 80 °C for 1 h to ensure full crosslinking of the PDMS. Similarly, a glass coverslip (170um) was applied on top of the PDMS membrane. The fabricated micro-resistors were measured to have an average electrical resistance of $R_{av} \approx 10$ kΩ.

Smartlens Electronics

The applied voltage is independently delivered to each SLs using an home-made electronic module, based on a PCA9685 12-bit Pulse Width Modulation (PWM) integrated circuit. The mean applied voltage is set by controlling the duty cycle of a high-frequency (1.5 KHz) square-modulated voltage (0 V – Vmax= 60 V). The module is driven by a Raspberry 4 computer via a Python script.

## Smartlens Characterization

The transmission spectrum of a single SL was measured by imaging its active area into a fibered spectrometer (Avantes, AvaSpec-ULS2048CL-EVO), and normalizing the measured spectrum by the one of the Halogen light source.

Phase responses were measured using quadriwave lateral shearing interferometry.[53] The SLs were illuminated with a halogen lamp, spectrally filtered (λ=605 nm, Δλ = 55 nm), and spatially filtered to provide a high spatial coherence (NAill<0.1). The SLs were imaged using a microscope objective (Olympus PlanN ×4, NA=0.1) on a high-resolution wavefront sensor (SID4, Phasics). The temperature maps were estimated from the phase images through a deconvolution procedure described in ref[54,55]. Briefly, the measured phase map is deconvolved by a phase Green's function $G_\varphi$ to estimate the power map $P$ dissipated by Joule effect. The temperature map is then obtained by convoluting this power map $P$ with the thermal Green's function $G_T = \frac{1}{4\pi\kappa r}$. See Supplementary section S2 for more detailed on the power consumption and the heat delivered by a SL. The SL vergence V was characterized using a parabola fit on the profile of the optical path difference: $\delta = \frac{1}{V} - \frac{V.x^2}{2}$, with x the distance.

## Setup specification

The excitation source used for fluorescence imaging is a Laser Quantum gem λ=473 nm. The beam is magnified to illuminates the active part of the DMD (Vialux V7001). The DMD is imaged to the IP with a magnification of 0.5 to fill the active area of the SL array (≈4×4mm), and reimaged through the object/sample plane of the microscope with an objective Olympus UMPlanFLN ×20, NA= 0.5. Excitation light is filtered by a dichroic mirror (STED laser beamsplitter zt 488 RDC, AHF) and a bandpass filter (520±28 nm Chroma). Samples are imaged on a camera Kinetix22 SCMOS. To fulfil Nyquist criteria, the PSF of the system is sampled by 2 pixels of the camera and 2 micromirrors of the DMD. The SL array is placed 7.2 mm before the IP (see Supplementary section S3 for more details). For the SIM illumination, DMD and camera were triggered by a function generator (Tektronix AFG 3102) to synchronize the SIM pattern and the recording. Camera acquisition and DMD control are made using a home-made

## Structured illumination

The method used to improve optical sectioning with the SIM modality was inspired by ref.[5] Three stripped illumination $U_1$, $U_2$ and $U_3$ were generated. The vectors unit $u_1, u_2, u_3$ related to the periodic pattern used for the illuminations $U_1$, $U_2$ and $U_3$ are defined as $u_1$=[0 0 1 1], $u_2$=[1 0 0 1] and $u_3$=[0 1 1 0], where the values 1 or 0 correspond to the light modulation. To maximize the optical sectioning, the "length" of periodic pattern was set to 16 DMD pixels (2 times the PSF diameter time the length of the vector u). The images $I_1$, $I_2$ and $I_3$ related to the illumination patterns $U_1$, $U_2$ and $U_3$ allows to reconstruct an image with improved optical sectioning using the following relation:

$$I_{SIM} = \sqrt{(I_1 - I_2)^2 + (I_1 - I_3)^2 + (I_2 - I_3)^2}.$$

Data processing

To calculate the ratio $\Delta F/F$, the fluorescent background signals F of the traces were calculated using a Fourier analysis of the temporal signal. $\Delta F/F$ was then calculated by subtracting the signal by F and dividing the result by F. Activity traces were obtained using Imagej/Fidji software and Matlab2023. Peaks from Figure 3d were detected from the traces Figure 3c using the following criteria: minimum peak distance=0.2 sec, minimum $\Delta F/F$=1.6, minimum peak width=0.1 sec, minimum peak prominence=60%.3D videos were made using Napari software.

**Data availability**

Due to the large volume of data acquired, the data supporting the findings of this study are available from the corresponding author upon reasonable request. Source data is provided with the manuscript, along with representative datasets.

## Acknowledgement

The project has received funding from the Agence Nationale de la Recherche and the Swiss National Funding Science Fundation (IHU FOReSIGHT ANR-18-IAHU-01, ANR-21-CE42–0023–01 PROFIT, SNF 205936 PROFIT), and the European Union's Horizon Europe research and innovation programme under grant agreement No.101063802. The vision Institute is affiliated to DIM C-BRAINS, funded by the Conseil Régional d'Ile-de-France. P.B. acknowledges support from Institut Universitaire de France. The authors thanks Raphael T. Steffen and Reda Berrada for preliminary experiments, Christophe Tourain for technical support as well as François Blot, Ruth Sims and Dimitrii Tanese for fruitful discussions.

**Author contribution** P.B, R.Q and G.T initiated the project. H.L.M. carried out the experiment, processed and interpreted the data, and developped the software for controlling the DMD and the camera. N.R. and J.G.G. designed and fabricated the ITO structures. C.L. and A.A. performed preliminary experiments. A.A. programmed the software for the control of the electrical module. G.F. injected and prepared the zebrafish larval samples, helped in the functional experiments and data interpretation. E.P generated the transgenic line, under the guidance of F.B. All authors participated to writing of the manuscript.

## Competing interests
The authors declare no conflicts of interest.